\documentclass[fleqn,usenatbib]{mnras}

\usepackage{graphicx}	
\usepackage{amsmath}	
\usepackage{amssymb}	
\usepackage{adjustbox}
\usepackage{threeparttable}
\usepackage{tabto}
\usepackage{float}
\usepackage{color}
\usepackage{graphics}
\usepackage{adjustbox}
\usepackage[usenames,dvipsnames]{pstricks}
\usepackage{indentfirst} 
\usepackage{natbib} 
\usepackage{threeparttable}
\usepackage{longtable}
\usepackage{multicol} 
\usepackage{multirow} 
\usepackage{epsfig}
\usepackage{pst-grad} 
\usepackage{pst-plot} 
\usepackage{pdflscape}
\usepackage{times} 

\newcommand{\Msun}{M$_{\odot}$}

\newcommand{\bse}{{\sc bse}}

\definecolor{smalt(darkpowderblue)}{rgb}{0.0, 0.2, 0.6}
\definecolor{forestgreen(traditional)}{rgb}{0.0, 0.5, 0.0}
\definecolor{red(traditional)}{rgb}{1.0, 0.15, 0.07}

\defcitealias{WW02}{WW02}
\defcitealias{LWW94}{LWW94}

\usepackage{etoolbox}
\makeatletter
\patchcmd\@combinedblfloats{\box\@outputbox}{\unvbox\@outputbox}{}{%
   \errmessage{\noexpand\@combinedblfloats could not be patched}%
}%
 \makeatother

\title[Are $B_{\rm WD}$ in CBs generated during CE evolution?]
{Are white dwarf magnetic fields 
in close binaries generated 
during common-envelope evolution?
}

\author[D. Belloni \& M. R. Schreiber]{
Diogo Belloni$^{1,2}$\thanks{diogo.belloni@inpe.br (DB)} and
Matthias R. Schreiber$^{2,3}$\thanks{matthias.schreiber@uv.cl (MRS)}\\
$^{1}$National Institute for Space Research, Av. dos Astronautas, 1758, 12227-010, S\~ao Jos\'e dos Campos, SP, Brazil\\
$^{2}$ Instituto de F{\'i}sica y Astronom{\'i}a, Universidad de Valpara{\'i}so, Av. Gran Breta{\~n}a 1111 Valpara{\'i}so, Chile \\
$^{3}$ Millenium Nucleus for Planet Formation, Universidad de Valpara{\'i}so, Valpara{\'i}so 2360102, Chile}

\date{Accepted XXX. Received YYY; in original form ZZZ}

\pubyear{2019}

\begin{document}
\label{firstpage}
\pagerange{\pageref{firstpage}--\pageref{lastpage}}
\maketitle

\begin{abstract}
Understanding the origin of the magnetic fields in white dwarfs (WDs) has been a puzzle for decades. A scenario that has gained considerable attention in the past years assumes that such magnetic fields are generated through a dynamo process during common-envelope evolution. We performed binary population models using an up-to-date version of the \bse~code to confront the predictions of this model with observational results. We found that this hypothesis can only explain the observed distribution of WD magnetic fields in polars and pre-polars and the low-temperature WDs in pre-polars if it is re-scaled to fit the observational data. Furthermore, in its present version, the model fails to explain the absence of young close detached WD+M-dwarf binaries harbouring hot magnetic WDs and predicts that the overwhelming majority of WDs in close binaries should be strongly magnetic, which is also in serious conflict with the observations. We conclude that either the common-envelope dynamo scenario needs to be substantially revised or a different mechanism is responsible for the generation of strong WD magnetic fields in close binaries.
\end{abstract}


\section{Introduction}

White dwarfs (WDs) in cataclysmic variables (CVs)
are more frequently magnetic and have, on average,
stronger magnetic fields than single WDs
\citep[e.g.][]{Ferrario_2015},
while the population of observed close detached 
WD+M-dwarf post-common-envelope binaries 
(PCEBs) is dominated by systems with 
negligible WD magnetic fields
\citep{Liebert_2015}.
Understanding these differences
may provide insight about magnetic 
field generation 
with implications beyond WD research.

In recent years, several hypothesis 
have been put forward to explain 
magnetic field generation in WDs. 
In the fossil field scenario
\citep[e.g.][]{Angel_1981},
it is assumed that
the magnetic flux is conserved 
during the WD formation and that
strongly magnetic Ap and Bp stars are 
the progenitors of magnetic WDs.  
However, \citet{Kawkaetal_2007} showed 
that the magnetic Ap and Bp stars 
cannot be the only progenitors of 
magnetic WDs as their birth rate is 
simply too small.  
In an alternative scenario, 
strong magnetic field generation  
occurs in the corona present 
in the outer layers of the
remnant of coalescing double WDs
\citep{Berro_2012}. 
This scenario, however, can only explain the 
large field strength of massive magnetic single
WDs but is not applicable to CVs or their detached progenitors. 
More recently, \citet{Isern_2017}
argued that when the WD
temperature is low enough and  
its interior crystallizes, 
a dynamo similar to those operating 
in main-sequence stars 
and planets can generate a magnetic
field. While this mechanism may work 
in both single WDs and WDs in close binaries, 
the field strengths predicted by 
\citet{Isern_2017}
are much smaller than those  
derived from observations of 
strongly magnetic WDs in CVs and
detached post-common-envelope 
binaries (PCEBs).

A hypothesis that gained significant 
attention during the last years and that 
has recently been claimed to 
fully explain the magnetic fields observed 
in WDs has been put forward by 
\citet{Tout_2008}. 
According to this scenario the 
high magnetic fields in WDs are generated 
by a dynamo created during the 
common-envelope (CE) evolution.
Based on this CE dynamo hypothesis,
\citet{Briggs_2018b} investigated 
the origin of WD magnetic fields in CVs
and claimed that this scenario 
can explain the observed characteristics 
of magnetic CVs which, if true, would 
provide considerable support for the 
CE dynamo hypothesis. 
However,
these authors only compared 
the predicted and observed 
WD magnetic field distribution 
of all close WD binaries
harbouring main-sequence stars.
While this is a first step in confronting 
the model with observations, a separate 
comparison of 
model predictions and observations 
for detached 
PCEBs and CVs provides crucial additional
constraints on the model.
This is particularly true because
the fraction of magnetic systems 
and
the underlying WD masses and orbital periods
are very different for both populations. 
Therefore, 
the question whether dynamo 
processes generated 
during CE evolution
can indeed 
explain occurrence rates and 
field strength of magnetic WDs in 
PCEBs and CVs remains unanswered.

We test here the CE dynamo hypothesis 
using binary population models 
of magnetic CVs performed with an updated 
version of the
\bse\,code, which includes state-of-the-art
prescriptions for the CE evolution and 
mass transfer stability. 
The new code furthermore takes into account the impact 
of the WD magnetic field on magnetic braking.
We compare the model predictions 
with the main observed properties of 
magnetic CVs and their progenitors, i.e.
(i) the WD magnetic field distribution
of magnetic CVs
\citep{Ferrario_2015};
(ii) the WD magnetic field,
the WD effective temperature 
and orbital period
distributions of pre-polars
\citep[][and references therein]
{Schwope_2009,Parsons_2013}; 
and
(iii) the relative numbers of  
magnetic WDs among 
close detached WD+M-dwarf PCEBs
\citep{Liebert_2015} 
and 
CVs
\citep{Pala_2019}.

\section{Binary Population Model}
\label{approach}

In order to test the origin of 
WD magnetic fields during CE evolution, 
we carried out binary population
synthesis with the 
\bse\,code 
\citep{Hurley_2002}, 
which has recently been modified and calibrated 
\footnote{\href{http://www.ifa.uv.cl/bse}{\texttt{http://www.ifa.uv.cl/bse}}} 
to carry out population synthesis 
of non-magnetic  
\citep{Belloni_2018b}
and magnetic CVs 
\citep{Belloni_2019b}.

\subsection{General assumptions}
\label{approach1}

The binary population simulations 
presented here are similar
to those shown in \citet{Belloni_2019b}.
In brief, we first generated 
an initial population 
of $2\times10^7$ binaries 
using the following 
initial distributions.
The primary mass was obtained from the 
canonical \citet{Kroupa_2001} initial 
mass function  
(i.e. with two stellar segments) 
in the range $[1,8]$ M$_\odot$;
the secondary was generated assuming
a uniform mass ratio distribution, 
where $M_2 \leq M_1$, and requesting 
that $M_2\geq0.07$; 
the semi-major axis was assumed to follow a 
log-uniform distribution in the 
range $[10^{-0.5},10^{4.5}]$ R$_\odot$
and the eccentricity to follow a 
thermal distribution in the range 
$[0,1]$.

We then evolved the generated binary star systems and 
selected those that 
start dynamically unstable mass 
transfer when the primary 
was on the first giant branch or the asymptotic 
giant branch. 
The critical mass ratio 
$q_c$ 
separating stable and unstable mass transfer 
adopted here
is based on the assumption of conservative mass transfer
and the condensed polytropic models 
by
\citet{HW1987}, 
i.e. {${q_c=0.362 + [3(1 - M_c/M_g)]^{-1}}$}, 
where 
$M_c$ is the giant core mass 
and 
$M_g$ is the giant mass.
Dynamically unstable mass transfer
gives rise to
CE evolution, which we modelled using
eqs. 69--77 of \citet{Hurley_2002}, taking into account
the upgrades described in appendix A of
\citet{Claeys_2014} related to the binding
energy parameter.
We considered three relatively 
small values for the CE efficiency 
$\alpha$ (0.1, 0.25 and 0.4),
assumed that no recombination energy 
contributes to the CE ejection
and computed the binding energy parameter
of each system based on the properties 
of the giant star. These assumptions have been shown to be  
reasonable in simulations of  
CVs and PCEBs
\citep[e.g.][]
{Zorotovic_2010,
Toonen_2013,
Camacho_2014,
Cojocaru_2017,
Belloni_2019a}.

For those binaries that survived CE evolution, 
we assumed standard angular momentum loss
prescriptions \citep[][]{Knigge_2011_OK}.
For the second phase of mass transfer, i.e. 
for the CV stage, we adopted the recently 
suggested empirical model for 
consequential angular momentum loss 
\citep[eCAML;][]{Schreiber_2016}. 
Observational evidence for this new model 
for CV evolution is growing. 
It is not only a good candidate to 
solve some long-standing problems 
related to CV evolution models, like 
the predicted large fraction of 
low-mass WDs in CVs, 
the predicted excess of 
short-period systems, 
and the overestimated space density 
\citep[][]{Schreiber_2016,Belloni_2018b,McAllister_2019}.
The eCAML idea also explains 
the existence of single low-mass WDs 
\citep{Zorotovic_2017}, 
the properties of detached CVs crossing
the orbital period gap
\citep{Zorotovic_2016}
and the characteristics of CVs in globular
clusters
\citep{Belloni_2019a}.

Additionally,
we do not consider CVs originating from a 
phase of thermal time-scale mass transfer,
since 
the \bse~code is unable to properly model
this phase 
and
observations show that only $\approx5$~per~cent 
of all CVs emerge from this channel
\citep{Pala_2019}.
We furthermore assume that in CVs the WD expels 
the accreted mass in repeated nova 
eruptions and, therefore, treat its mass as 
constant during CV evolution.
All other stellar/binary evolution parameters 
not mentioned
here are set as in \citet{Hurley_2002}.

\subsection{Assumptions related to the WD magnetic fields}
\label{approach2}

Concerning the influence of the WD
magnetic field on CV evolution, 
we adopted the  
reduced magnetic braking model 
proposed by 
\citet*{LWW94}
and developed further 
by \citet{WW02}. 
This approach can reasonably well
explain the observed properties 
of polars \citep{Belloni_2019b} if the 
WD magnetic field strength distribution 
is assumed to be the observed one.

In order to test the scenario of 
magnetic field generation during CE 
evolution, we changed our code and,
instead of using the observed distribution,
we determined the WD magnetic field 
strength $B_{\rm WD}$ 
in each PCEB using the
formula provided by \citet{Briggs_2018a}, 
i.e.

\begin{equation}
B_{\rm WD} = 1.35\times 10^{10}
\left(
\frac{\Omega}{\Omega_{\rm crit}}
\right)~~\mbox{G}~~,
\label{BWD}
\end{equation}
\

\noindent
where 
$\Omega$ is the orbital angular velocity 
just after the CE evolution given by 

\begin{equation}
\Omega
= 
\frac{2\,\pi}{P_{\rm orb}}
~~~~{\rm yr}^{-1}~~,
\label{OMEGAORB}
\end{equation}
\

\noindent
with $P_{\rm orb}$ being the orbital
period just after the CE evolution
and
$\Omega_{\rm crit}$ the break-up 
angular velocity of the WD given by

\begin{equation}
\Omega_{\rm crit} 
= 
\sqrt{\frac{GM_{\rm WD}}{R_{\rm WD}^3}}
=
2\,\pi\,
\sqrt{
  \left(
    \frac{M_{\rm WD}}{{\rm M}_\odot}
  \right)
  \left(
    \frac{R_{\rm WD}}{\rm AU}
  \right)^{-3}
}
~~~~{\rm yr}^{-1}~~,
\label{OMEGAWD}
\end{equation}
\

\noindent
where $M_{\rm WD}$ and $R_{\rm WD}$
are the WD mass and radius,
respectively.
We additionally assume that 
$B_{\rm WD}$ is constant during 
PCEB and CV evolution.

As in \citet{Briggs_2018b}, we assumed that 
magnetic fields are not generated in 
any CE event.
Systems in which
either 
(i) the giant has a non-degenerate core,
or 
(ii) the proto-WD experiences further nuclear burning 
are assumed to form PCEBs with non-magnetic WDs. 
The reasons for these additional conditions
are that in a non-degenerate core a magnetic field 
cannot be maintained in a frozen-in state
and that nuclear burning in the proto-WD naturally
induces convection which would destroy any frozen-in 
magnetic field. 

The selection of CE events with a degenerate core (point i) 
is implemented using 
the critical zero-age main-sequence
mass ($M_{\rm HeF}$) which 
separates low-mass stars that develop
a degenerate core on the first giant branch
from more massive ones which only develop a degenerate core 
on the asymptotic giant branch.
We adopted here the standard value for
$M_{\rm HeF}$ from \bse, i.e. 
$1.995$~\Msun, for solar metallicity
\citep[][their eq.~2]{Hurley_2000}.
CE events with giants that did not develop a 
degenerate core then occur 
either if a giant with initial mass smaller than
$M_{\rm HeF}$ fills its Roche lobe 
during the sub-giant phase,
or if a giant with initial mass larger than
$M_{\rm HeF}$
fills its Roche lobe during either 
the sub-giant or the first giant branch phase.
We note that CE events with giants having a
non-degenerate core are likely to result in mergers,
since, in these cases, the initial 
orbital separation must be relatively small so that the 
binary orbital energy is typically not large enough 
to prevent the binary coalescence.

The second condition for the existence of magnetic fields, 
i.e. no post-CE nuclear burning (point ii), 
can be violated
if the giant progenitor was
relatively close to the tip of
the first giant branch at the beginning 
of the dynamically unstable mass transfer that generated CE evolution. 
In this case the degenerate core may ignite following CE evolution 
which results in a hot 
B-type subdwarf. These naked
helium-burning stars cannot maintain 
the magnetic field generated during 
the CE evolution.
As in \citet{Zorotovic_2013},
we select these systems following 
\citet{Han_2002}. These authors performed 
a comprehensive series of stellar evolution
calculations, assuming that mass ejection 
during a CE event takes place on much shorter
time-scales than in single giant star
evolution.
These models provide the 
minimum core mass as a function of the initial
mass, above which the core will 
still ignite helium after the CE ejection. 
These minimum core masses and initial masses are listed in their table 1.
We here adopted their value for solar metallicity,
with stellar wind and convective overshooting, and
linearly interpolated their grid to determine
the minimum proto-WD mass needed to trigger 
helium burning as a function of the initial mass.

While these two 
additional criteria may have a minor impact on the predicted PCEB population, 
they clearly have no impact for the predicted CV populations, 
since they are only applicable 
for progenitor systems of low-mass WDs 
(${\lesssim0.5}$~\Msun).
Observed CV WD masses, however,  
are always ${\gtrsim0.5}$~\Msun~
\citep{McAllister_2019,Zorotovic_2011}
and the eCAML model 
\citep{Schreiber_2016} adopted in our simulations 
always provides CV WD masses
${\gtrsim0.5}$~\Msun, consistent with observations.

When comparing our model predictions to observed 
populations we considered only CVs with 
donor masses greater than {$0.05$\,\Msun}
and PCEBs having secondary 
masses smaller than {$0.6$\,\Msun}.
In other words, 
we neglect period bouncers
and concentrate on systems with 
M-dwarf companions in PCEBs.
The reason for both these limits 
are potential strong observational 
biases in the observed samples. 
Period bouncers are hard to find 
because of their extremely low 
mass transfer rates.  
PCEBs with secondary stars 
earlier than spectral type M 
are often overlooked as the optical 
emission is entirely dominated by the 
main-sequence companion which makes it 
difficult to detect the WD component. 
Finally, we 
define a limit of $B_{\rm WD}=1$\,MG to 
separate magnetic and and non-magnetic
systems (either PCEBs or CVs).
This strict limit is somewhat arbitrary 
but roughly reflects the minimum 
field strengths that have been measured
for WDs in PCEBs and CVs.

\section{Confronting the model with observations}

If  the model proposed by 
\citet{Briggs_2018b} was correct, the
resulting predictions for magnetic WDs 
in all WD binaries
should resemble their observed properties.   
The ideal systems to carry out this
comparison between model predictions 
and observations are the large 
populations of 
detached WD+M-dwarf PCEBs
and
CVs.

\subsection{Post-Common-Envelope Binaries}
\label{resultsPCEB}

Observations clearly show that the number 
of magnetic systems among PCEBs is small. 
The population of observed PCEBs is dominated 
by systems with negligible $B_{\rm WD}$
\citep[][and references therein]{Liebert_2015}.
Only ten PCEBs with strongly magnetic WDs, 
so called \textit{pre-polars}, 
have been identified so far
\citep[][and references therein]
{Schwope_2009,Parsons_2013}.
Given that the \emph{Sloan Digital Sky Survey}
(SDSS) alone has discovered
several hundred PCEBs
\citep{schreiberetal10-1}, 
we can safely state that the observed 
fraction of magnetic PCEBs is 
well below ten per cent.
All the magnetic WDs in PCEBs are relatively 
cool $(T\lesssim10^4$~K) and they seem to be 
rather close to Roche-lobe filling 
as the WDs accrete from the wind of 
their M-dwarf secondaries via a magnetic 
siphon.
The resulting mass transfer rates are very 
low $(\sim10^{-14}$~\Msun~yr$^{-1})$.
None of the magnetic WDs in 
detached systems is a He-core WD
\citep[][]{Mansergas_2011}.

Our binary population models 
predict that the overwhelming majority 
of PCEBs have orbital periods shorter than
$\sim5$\,days and in general small 
$M_{\rm WD}$ ($\sim0.45-0.55$\,\Msun). 
Both these predictions are in good 
agreement with observations of PCEBs
\citep[e.g.][]
{schreiberetal10-1,
Zorotovic_2010,
Zorotovic_2011,
Moran_2011}.
However, if combined with Eq.\,\ref{BWD}, 
these otherwise reasonable predictions 
produce an extremely high fraction of 
magnetic PCEBs.
The post hot subdwarfs 
binaries that are assumed to be non-mangetic 
make up a small fraction of the
PCEB population
($\lesssim18$~per~cent), 
which is consistent with
\citet{Zorotovic_2013},
who found a fraction of 
$\sim16$~per~cent.
This relatively small fraction is a direct consequence of
the minimum WD mass needed to trigger further
nuclear evolution, which results in 
WD masses lying in a very narrow range 
($\sim0.38-0.45$~\Msun).
After removing these core-helium burning
proto-WDs, Eq.\,\ref{BWD} provides that
about $60-90$ per cent 
of all PCEBs are magnetic,
depending on the CE efficiency.
In addition, using Eq.\,\ref{BWD},  
the model predicts that most 
systems with He-core WDs are 
magnetic, with $B_{\rm WD}$ ranging 
from $\sim1$ to $\sim100$\,MG.
The predicted large fraction 
of magnetic systems and especially the large 
fraction of magnetic He-core WDs in PCEBs
predicted by Eq.~\ref{BWD}
are in strong disagreement with the 
observations.

\begin{figure}
   \begin{center}
    \includegraphics[width=0.995\linewidth]
    {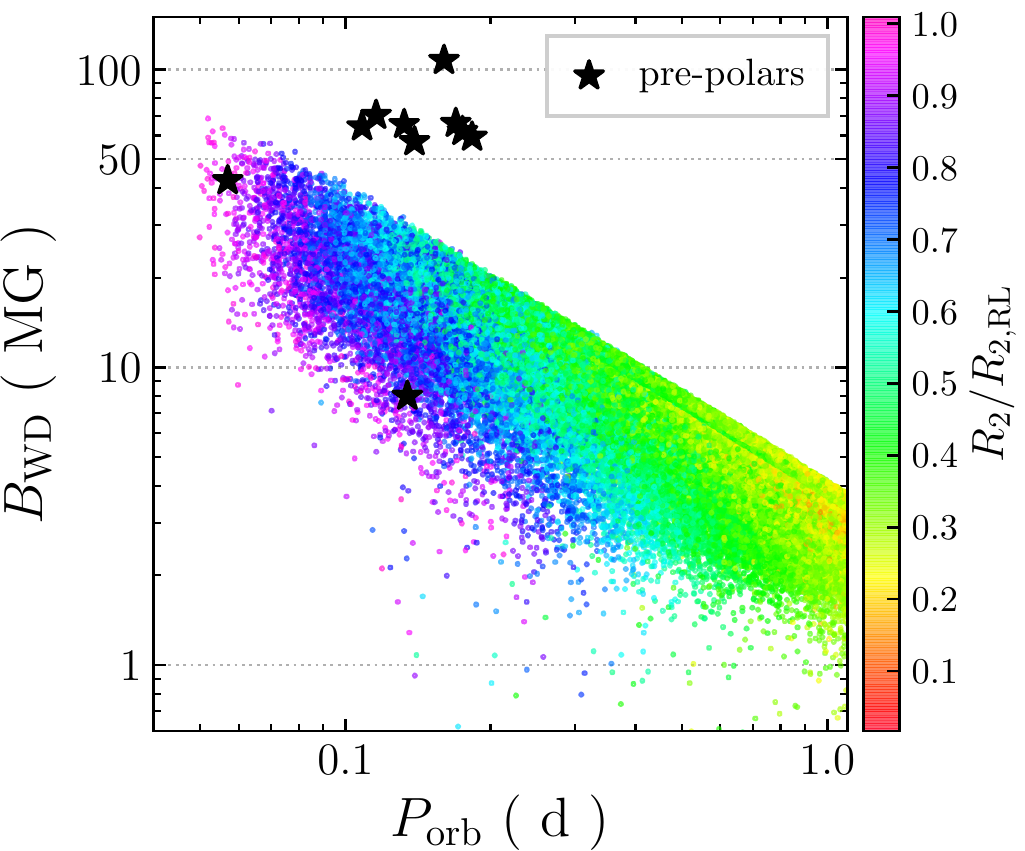}
    \end{center}
  \caption{Distribution of predicted
  PCEBs and observed pre-polars
  in the plane orbital period 
  ($P_{\rm orb}$)
  versus WD magnetic field
  ($B_{\rm WD}$).
  We show the model for the CE
  efficiency $\alpha=0.25$.
  Colours indicate the secondary
  Roche-lobe overfilling factor
  for simulated PCEBs.
  Observed measurements are from
  \citet[][and references therein]
  {Schwope_2009} and
  \citet{Parsons_2013}.
  Notice that, with the exception
  of two systems, predicted values
  for $B_{\rm WD}$ are not strong 
  enough to explain observed values 
  among pre-polars, provided their
  orbital periods.
  }
  \label{Fig03}
\end{figure}

The fraction of predicted magnetic WDs 
and its dependence on $M_{\rm WD}$ 
are not the only
predictions of Eq.~\ref{BWD} that can be 
confronted with observations.
In Fig.~\ref{Fig03}, we show 
$B_{\rm WD}$
as a function of orbital period 
for the simulated PCEBs 
(assuming a CE efficiency of 0.25) 
and the observed pre-polars. 
Apparently, with the exception
of two pre-polars, the predicted 
$B_{\rm WD}$ are significantly 
below the observed values, which 
cluster around $60-70$~MG.
Thus,  
despite predicting a far too large 
fraction of magnetic systems among 
PCEBs, Eq.~\ref{BWD} predicts relatively
weak WD magnetic fields and cannot explain 
the field strength of most 
(eight out of ten) pre-polars.

\begin{figure}
   \begin{center}
    \includegraphics[width=0.975\linewidth]
    {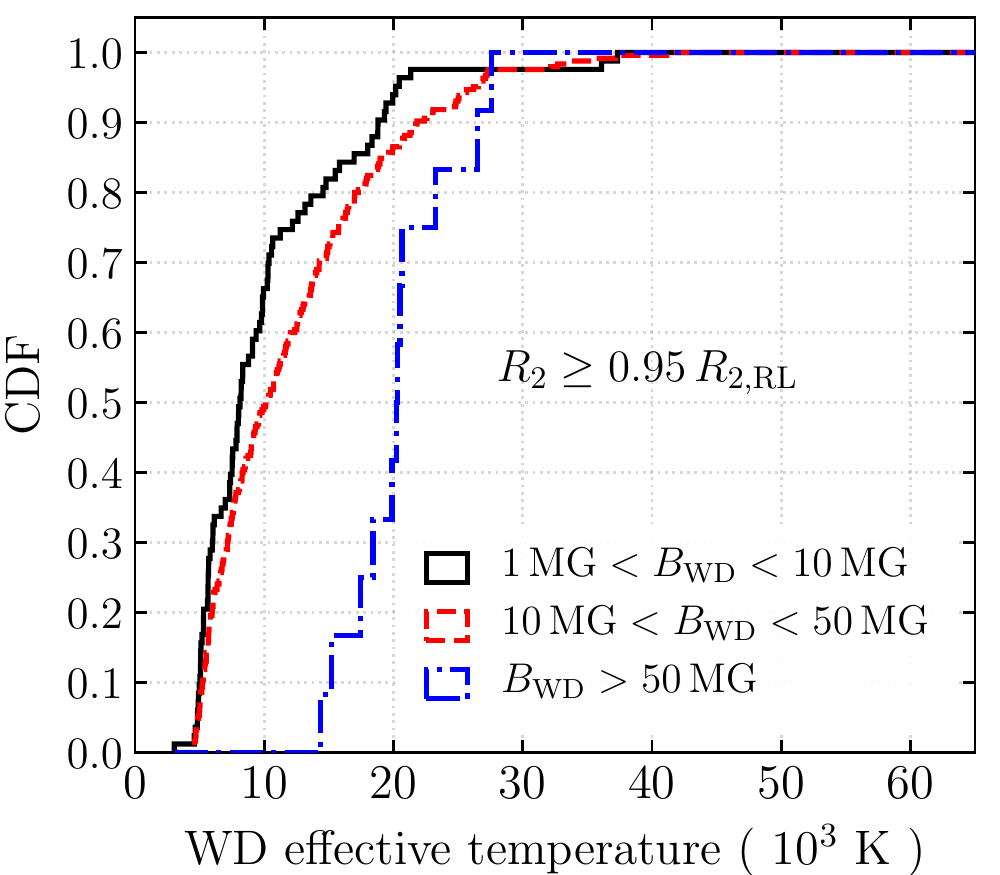}
    \end{center}
  \caption{Cumulative distribution 
  function of the WD effective 
  temperature in PCEBs whose secondaries
  fill at least $\sim95$ per cent of 
  their Roche lobe. 
  We show the case for the CE
  efficiency $\alpha=0.25$ and
  separate the population
  according to the strength
  of $B_{\rm WD}$, in units
  of MG.
  Notice that 
  {${\approx50}$}~per~cent 
  of systems with
  $10$~MG~$<B_{\rm WD}<$~$50$~MG
  have WD effective
  temperature 
  {{${\gtrsim10\,000}$}~K}.
  In addition, basically all WDs
  having $B_{\rm WD}>50$~MG are
  hotter than {{${15\,000}$}~K}.
  }
  \label{Fig02}
\end{figure}

We continue the comparison with observations
by addressing now the
low WD effective temperatures of 
the observed pre-polars. 
As nine 
pre-polars have 
secondaries that are very 
close to filling their Roche-lobes,
in order to properly compare with
observations, we selected
simulated PCEBs in which the
secondary is filling at least
$\sim95$ per cent of their Roche lobe,
i.e. {${R_2\geq0.95\,R_{2,{\rm RL}}}$}.
For each system, provided $M_{\rm WD}$
and age, we determined its effective
temperature by interpolating grids
of hydrogen-rich atmosphere WDs.
For He-core WDs 
{({$M_{\rm WD}\lesssim0.5$\,\Msun})},
we used the cooling tracks provided by 
\citet{Panei_2007};
for CO-core WDs 
{({${0.5\lesssim\,M_{\rm WD}\lesssim1.05}$})},
we used the cooling sequences 
of \citet{Renedo_2010} and;
for ONe-core WDs 
{({${M_{\rm WD}\gtrsim1.05}$\,\Msun})},
we used the evolutionary sequences 
of \citet{Althaus_2007}.

In Fig.~\ref{Fig02},
we show the resulting 
WD effective temperature
distributions, separated according
to the strength of $B_{\rm WD}$.
Virtually all systems with
$B_{\rm WD}$ stronger than
$\sim50$~MG contain WDs hotter than
$\sim15\,000$~K. 
This is in contradiction 
to the observations, 
as the WDs in eight 
pre-polars have fields stronger 
than $50$~MG and are colder
than $\sim10\,000$~K.
On the other hand,
the WD temperature
in the two pre-polars with fields weaker 
than $\sim50$\,MG can be explained by
the model as roughly half of the
simulated systems with fields between
$10$ and $50$\,MG have WDs
cooler than $10\,000$~K.

The general disagreement between 
predicted and observed WD temperatures of pre-polars 
with secondaries close to filling their Roche-lobe
is again a direct
consequence of Eq.~\ref{BWD}. 
In order to have $B_{\rm WD}$ stronger
than $\sim50$\,MG, the orbital
periods after CE evolution need 
to be very short.
This implies that such systems
will be closest to the 
CV phase after emerging from the CE 
evolution, and will consequently  
be the youngest and host the hottest WDs, when
the secondary is getting close to filling 
its Roche-lobe.

It furthermore appears difficult
to explain the identified 
discrepancy as an observational bias 
because current surveys, 
such as the SDSS,
efficiently detect WD+M-dwarf binaries 
with WD effective temperatures 
from $\sim7\,500$ 
to $\sim57\,000$\,K
\citep[e.g.][]{Zorotovic_2011}.
If CE evolution was responsible for
the magnetic field generation, 
one would expect large numbers of hot WDs
with strong $B_{\rm WD}$.
These hot magnetic WDs would clearly 
be detectable as being magnetic in 
surveys such as SDSS,
via the detection of Zeeman splittings 
from the surface of WDs with 
$B_{\rm WD}\gtrsim1$~MG
\citep{Kepler_2013}.
The M-dwarf companions do not 
significantly affect the WD spectrum
for WD temperatures exceeding 
$\sim25\,000$\,K 
and magnetic single WDs with such 
temperatures have been identified
\citep{Ferrario_2015}.
Therefore, 
the fact that not a single magnetic PCEB with a 
hot WD is known further 
suggests
that
the idea of 
generating $B_{\rm WD}$ during 
CE evolution, 
in its current form,
is in disagreement with the observations.

\subsection{Cataclysmic Variables}
\label{resultsCV}

One of the easiest and therefore most 
precise measurement available for CV 
populations is the fraction of magnetic 
systems.
A recent detailed study of CVs within 
150\,pc provided a measured value for 
the fraction of magnetic CVs of 
$\lesssim33$\,per cent
\citep[][and references therein]
{Pala_2019}. 
Our binary population model, however, 
predicts a much large fraction 
of at least $94$ per cent of all 
CVs being magnetic.
This large predicted fraction and the 
resulting huge discrepancy between 
theory and observations is a simple result 
of combining Eq.\,\ref{BWD} with realistic 
binary population models of CVs. 

\begin{figure}
   \begin{center}
    \includegraphics[width=0.975\linewidth]
    {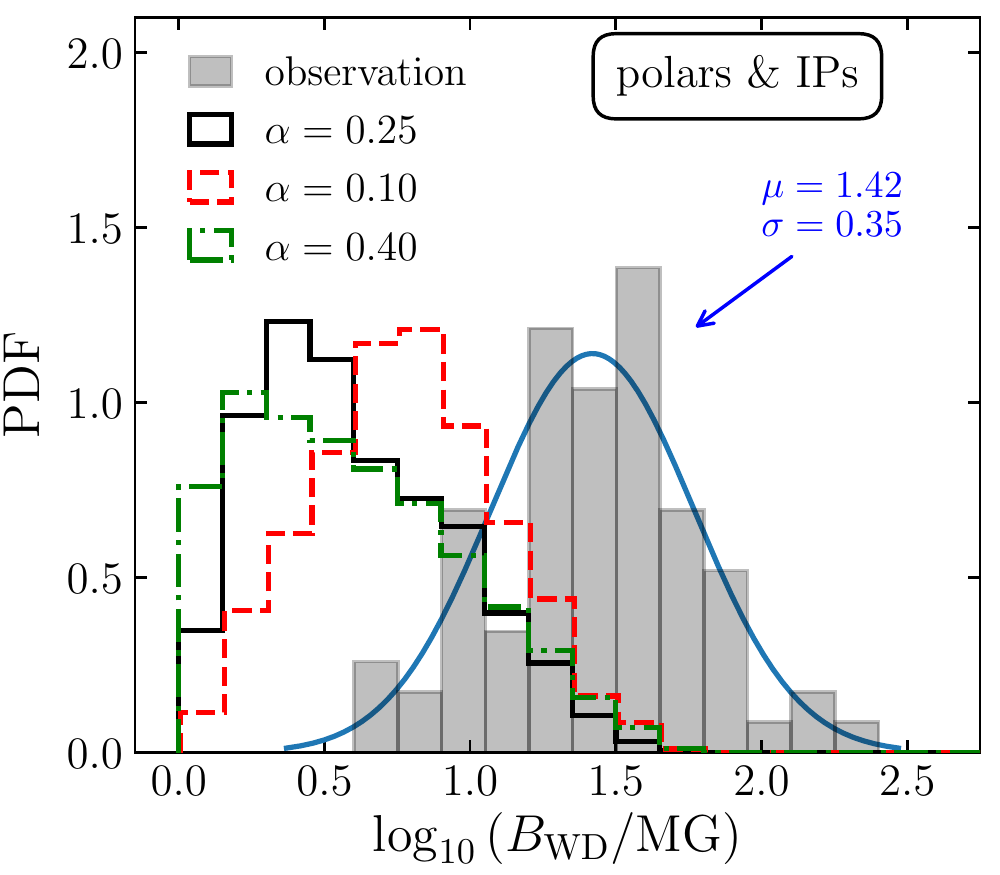}
    \end{center}
  \caption{
  Predicted and observed
  WD magnetic field strength $(B_{\rm WD})$
  distributions of magnetic CVs.
  Observed measurements 
  are from \citet{Ferrario_2015} and
  best-fitting Gaussian to the observed
  distribution is also shown.
  Note that predicted 
  distributions
  disagree with the 
  observational, irrespective
  of the CE efficiency $\alpha$.
  }
  \label{Fig01}
\end{figure}

The second observable we can compare 
with model predictions is the 
$B_{\rm WD}$ strength.
The observed distribution of 
magnetic CVs contains 
77 polars and intermediate
polars with measured $B_{\rm WD}$
\citep[][their tables 2 and 3]{Ferrario_2015}, 
peaks at 
$\log_{10}\left(B_{\rm WD}/{\rm MG}\right)\sim1.42$, 
and has a standard deviation
of $\sim0.35$.
In Fig.~\ref{Fig01},
we compare this distribution 
with the model predictions.  
The predicted distributions, 
according to Eq.~\ref{BWD},
contain much more low-field systems
than in the observed distribution, 
regardless of the CE efficiency
$\alpha$.
In particular, predicted $B_{\rm WD}$ 
are always weaker than $60$\,MG, 
which is 
below the values
measured for high-field polars. 
Were the observed $B_{\rm WD}$ 
be as low as predicted by 
Eq.~\ref{BWD}, intermediate polars
would 
dominate over polars,
which is not what observations show.
Among the predicted magnetic CVs
only $\sim25-30$ per cent are polars
({${B_{\rm WD}\gtrsim10}$\,MG})
while observations 
show that 
polars are more common than intermediate 
polars by a factor of $\sim2-4$
\citep[][]{Pretorius_2013,Pala_2019}.

While the observed $B_{\rm WD}$ distribution 
is likely somewhat biased and not necessarily 
representative for the intrinsic
population in the Galaxy, it is clear
from Fig.~\ref{Fig01} that Eq.~\ref{BWD}
does not provide $B_{\rm WD}$ strong
enough to explain the large fraction
of observed systems with 
$B_{\rm WD}\gtrsim60$\,MG.
Therefore, it seems unlikely 
that the $B_{\rm WD}$ distribution
of the intrinsic Galactic population of 
magnetic CVs, especially polars,
can be explained with the 
hypothesis of magnetic field 
formation during CE evolution 
in the way proposed by 
\citet{Briggs_2018a}.

\section{Why do our results differ from those of 
Briggs et al. ?}

While \citet{Briggs_2018b} claimed to find
agreement between model predictions and observations
when generating WD magnetic fields during CE evolution, 
we found here disagreement.
These different conclusions are obtained 
because of three reasons.

The first reason for the different 
results obtained 
by \citet{Briggs_2018b} and in this work 
lies in us 
addressing the full properties of
the few known pre-polars.
While \citet{Briggs_2018b} 
speculated that 
the initial rapid cooling of WDs could
explain the low WD effective temperatures
in these systems,
we compared here the 
WD effective temperatures, 
orbital periods, and 
$B_{\rm WD}$ of the observed pre-polars 
with our model predictions.

The second reason is connected with the 
fraction of magnetic systems
among the predicted populations. 
While this fraction was 
not computed by \citet{Briggs_2018b}, 
we found that Eq.~\ref{BWD}
predicts unrealistically high fractions of
both magnetic detached WD+M-dwarf PCEBs and CVs.

The third reason is that different 
binary population codes were used. 
Unlike \citet{Briggs_2018b}, we 
used here a model that includes the 
evolution of magnetic CVs 
\citep{Belloni_2019b}
and recent revisions 
suggested by \citet{Schreiber_2016}. 
The latter is crucial for comparing CV 
populations as only the revised model 
brings into 
agreement the predicted and observed 
orbital period distributions, 
WD mass distributions, and 
space densities.

\section{How to progress with the common-envelope 
dynamo scenario?}

So far, we have shown that the CE dynamo model,
as proposed by \citet{Briggs_2018a}, 
cannot explain the observations of 
magnetic WDs in close binaries.  
The main problems of the current formulation are that the model
\begin{enumerate}
    \item\noindent predicts WD magnetic fields too weak to explain those derived from observations of polars and pre-polars; 
    \vspace{0.15cm}
    \item\noindent predicts that pre-polars close to filling their Roche-lobe should mostly contain hot WDs, while observed ones are all cold; 
    \vspace{0.15cm}
    \item\noindent does not explain the lack of hot strongly magnetized WDs in the observed population of PCEBs;
    \vspace{0.15cm}
    \item\noindent predicts that most close WD binaries should harbour WDs with strong magnetic fields, which is inconsistent with measured fractions.
\end{enumerate}

In what follows we discuss whether plausible revisions of the
model exist that might bring into agreement theoretical predictions and observations.

\subsection{The field strength problem and the WD temperatures in pre-polars} 

The problem with the predicted field strength being far lower than the observed 
ones has a relatively straightforward solution. 
Given the simplicity of the model, one could just adapt 
the multiplicative factor 
({${B_0=1.35\times10^{10}}~{\rm G}$}) 
in Eq.~\ref{BWD}.  
Changing the value of this factor does not alter 
the shape of the distribution but only shifts 
the predicted field strengths 
to larger or smaller values.  
Therefore, increasing 
$B_0$ could easily bring into agreement 
the predicted and observed field strength distributions
shown in Figs.~\ref{Fig03}~and~\ref{Fig01}.

It also appears that the low temperatures
of the observed WDs in pre-polars
could at least be partly explained 
by increasing $B_0$. 
The fraction of cool WDs in systems close 
to filling their Roche-lobe in 
Fig.~\ref{Fig02} would significantly
increase for the field strengths 
($\sim50-70$~MG) of observed pre-polars.

Changing $B_0$ can therefore 
most likely fix the
field strength problems in close binaries.  
As the original value for $B_0$ has been obtained 
from fitting the magnetic field strength of 
single high field WDs assuming they are 
the outcome of a merger process during 
the CE evolution, this would imply that two different values of $B_0$  
would be required. 
Having different values 
for $B_0$ for binaries and single WDs 
would affect the general validity of 
Eq.~\ref{BWD} aimed for by 
\citet[][]{Briggs_2018a,Briggs_2018b}. 
However, given how simplistic the 
proposed model is, it might not be 
surprising that different values 
of $B_0$ are required for different 
types of objects.

\subsection{The missing young magnetic 
post-common-envelope binaries}

In order to explain the absence of young and hot 
PCEBs harbouring strongly magnetised WDs in observed samples,
an additional mechanism 
needs to be added to the model. 
If the WD magnetic fields are 
generated during CE evolution, 
their appearance in observed samples of 
close binaries must be delayed for
{${\sim0.5-1.5}$~Gyr} 
(the typical cooling age of WDs 
with effective temperatures
of {${\sim10\,000}$~K}).

One possibility to decrease the magnetic field strength for young PCEBs would be to 
assume that the fields are buried similarly to those of the weakly magnetised neutron 
stars in millisecond pulsars \citep{Romani1990}. 
For such a scenario to work, some material of the CE
must remain bound to the system and falls back onto the WD. 
Indeed, it has 
been claimed that up to
${\sim1-10}$~per~cent of the envelope 
material might 
remain bound to the binary following CE evolution \citep{Kashi+Soker2011}.
Additionally,
\citet{Zhang_2009} showed that {${\sim0.1-0.2}$~\Msun}~of accreted material 
are required to bury a strong WD magnetic field in CVs.
Thus, to fully bury
the generated magnetic field, virtually all the remaining material must be accreted 
by the WD. 
Furthermore, the magnetic field would need 
to be buried for a very long time, i.e. 
{${\sim0.5-1.5}$~Gyr} 
in a detached binary, 
i.e. without further accretion. 

For comparison, in the case of neutron stars,
the time-scale needed for 
magnetic fields buried by a 
post-supernova episode of 
hypercritical 
accretion 
\citep[e.g.][]
{Chevalier_1989,Geppert_1999,Bernal_2010}
to diffuse back to the surface 
is of the order of 
{${\sim10^3-10^4}$~yr}
\citep{Wynn_2011}. 
Thus, a successful model must explain why 
the WD magnetic fields generated during CE evolution are 
buried for time-scales several orders of magnitude longer than those in neutron stars.

Detailed and dedicated 
theoretical investigations of burying fields of magnetic WDs following CE evolution 
are required to further evaluate this possibility. 
Based on such detailed investigations one could hope 
to confront a quantitative description of 
the burying mechanism for WDs 
with observations.

\subsection{The fraction of magnetic systems}

With respect to the last 
problem of the CE dynamo model,
it is not obvious how the
fraction of magnetic systems predicted by the CE dynamo model 
could be decreased. It is clear that one would need to find a more 
complex dependency of magnetic field generation on the binary/CE parameters, so far not considered in the
model. 
In order to reproduce the observed fraction
of magnetic systems in close WD binaries,
such a more complex form of Eq.~\ref{BWD} 
should permit strong WD magnetic field 
generation \textit{only}
in a very small subset of CE events.

Typical CE dynamo models for the generation of 
magnetic fields assume that the dynamo processes 
are driven by shear due to differential rotation 
in the envelope 
\citep{Regos_1995,Potter_2010}, 
in an accretion disc
\citep{Nordhaus_2011},
or 
in the hot outer layers of the 
degenerate core
\citep{Wickramasinghe_2014}.
According to these models, 
several properties play an important role 
for amplifying and maintaining 
the magnetic field.
Among them are 
the differential rotation,
the CE mass, radius and density, 
the total mass of the binary, 
the total energy generated inside the CE,
the orbital energy and angular momentum,
the radius of the convective zone, 
i.e. the interface between 
the convective and radiative regions, 
the thickness of the convective zone,
as well as 
the life-time of the dynamo activity.

However, which of these parameters are the most
important ones involved in 
CE evolution and the claimed dynamo process  
is currently unclear. 
No numerical 
approach capable of fully addressing the
physical mechanisms and time-scales 
involved in CE evolution  
has been suggested yet
\citep[e.g.][]{Ivanova_REVIEW}.
It is furthermore not at all clear under which conditions
the magnetic fields produced from such dynamos
are persistent 
\citep[e.g.][]{Potter_2010}
and likely to 
reach (or be generated on)
the WD surface with sufficient 
strength to 
explain observations of 
WDs in close binaries
\citep[e.g.][]{Ohlmann_2016}.
Therefore, it remains uncertain
whether a more complex version of 
Eq.~\ref{BWD} might be able to 
significantly reduce the predicted fraction 
of magnetic post-CE systems and, 
at the same time,
provide sufficiently strong magnetic field 
in some systems
to explain the WD magnetic fields observed in pre-polars and CVs.

\section{Summary and Conclusions}

Explaining the origin of magnetic 
fields in WDs has been a challenge 
for decades.
A handful of mechanisms have been proposed, but none of them is yet considered to be 
fully convincing. 
One scenario that has gained some
attention in the past years is
the model in which the WD magnetic
field is generated via a dynamo
process during common-envelope
evolution.
We examined whether 
such a scenario could explain 
the observed fraction of magnetic 
cataclysmic variables,
the observed distribution
of WD magnetic fields in 
polars and pre-polars, 
the incidence of cool WDs
amongst pre-polars and the
paucity of detached WD+M-dwarf
post-common-envelope binaries
harbouring magnetic WDs.

By performing binary population synthesis 
with a state-of-the-art version of the
\bse~code, 
we found that this scenario needs to be 
re-scaled 
to explain the WD magnetic field distributions of 
polars and pre-polars as well as the
observed low temperatures of the WDs in pre-polars.
In order to explain the absence of young 
detached WD+M-dwarf post-common-envelope binaries
harbouring hot and magnetic WDs, a more severe 
revision of the model would be required. 
Somehow the magnetic fields generated during 
common-envelope evolution need to be 
buried for {${\sim0.5-1.5}$~Gyr}. 
While this cannot be excluded, there is currently
no detailed physical description for a mechanism 
able to bury the magnetic field for such a long 
time.
Finally, even with these modifications,
the common-envelope dynamo scenario would still produce an
unrealistically high fraction of systems
containing magnetic WDs among cataclysmic variables, which indicates
that the model is currently too simplistic,
and a more complex dependency on the 
binary/common-envelope parameters is needed
so that negligible fields are generated in 
most common-envelope events.

We conclude that the current model 
is facing serious challenges 
and needs to be substantially 
improved to account for the 
observed properties of magnetic 
cataclysmic variables 
and 
their detached progenitors. 
Alternatively, another process
might be responsible for 
the WD magnetic field 
generation in close WD binaries.

\section*{Acknowledgements}

We would like to thank an anonymous referee for the comments and suggestions that helped to improve this manuscript.
We also thank Lilia Ferrario for her feedback on the first version of this paper.
We thank MCTIC/FINEP (CT-INFRA grant 0112052700) and the Embrace Space Weather Program for the computing facilities at the National Institute for Space Research, Brazil.
DB was supported by the grants \#2017/14289-3 and \#2018/23562-8, S\~ao Paulo Research Foundation (FAPESP).
MRS acknowledges financial support from FONDECYT grant number 1181404.


\bibliographystyle{mnras}
\bibliography{references} 


\bsp	
\label{lastpage}
\end{document}